\documentclass[twocolumn,preprintnumbers,amsmath,amssymb,floatfix]{revtex4}

\usepackage{graphicx}
\usepackage{dcolumn}
\usepackage{bm}

\newcommand{\beq}{\begin{equation}}
\newcommand{\eeq}{\end{equation}}
\newcommand{\bey}{\begin{eqnarray}}
\newcommand{\eey}{\end{eqnarray}}

\begin{document}

\title{Energy Distribution for Non-commutative Radiating Schwarzschild Black
Holes}
\author{I. Radinschi$^{1}$, F. Rahaman$^{2}$ and U. F. Mondal$^{3}$ \\
$^{1}$Department of Physics, \textquotedblleft Gh. Asachi\textquotedblright\
Technical University, Iasi, 700050, Romania\\
E-mail:  radinschi@yahoo.com\\
$^{2}$Department of Mathematics, Jadavpur University, India\\
E-mail:
rahaman@iucaa.ernet.in\\
$^{3}$ Department of Mathematics Behala College,  Parnasree,
Kolkata 700060 , India\\
E-mail:umarfarooquemondal@ymail.com}

\date{\today}

\begin{abstract}
The aim of this article is the calculation of the energy-momentum
for a non-commutative radiating Schwarzschild black hole in order
to obtain the expressions for energy. We make the calculations
with the Einstein and M\o ller prescriptions. We show that the
expressions for energy in both the prescriptions depend on the
mass $M$ ,  $\theta $ parameter  and radial coordinate. We make
some comparisons between the results. Our results show that
 the Einstein prescription is
a more powerful concept than the M\o ller prescription.
\end{abstract}

\maketitle

\section{Introduction}

Among the many interesting issues of General Relativity one of the
most important is the energy-momentum localization. To solve the
problem of energy-momentum localization means to develop an unique
mathematical formula for energy density. Nowadays, in General
Relativity there are some well-known tools for the calculations of
the energy-momentum like superenergy tensors [1], quasi-local
expressions [2], the energy-momentum complexes of Einstein [3],
Landau-Lifshitz [4], Papapetrou [5], Bergmann-Thomson [6],
Weinberg [7], Qadir-Sharif [8] and M\o ller [9] and the
tele-parallel theory of gravitation [10]. The tele-parallel theory
of gravitation [10] presents the advantage that the calculations
can be performed in such a manner that the problem of the
coordinate dependence can be avoided.
\\
\\ The pseudotensorial definitions [3]-[9] have been used by many authors and have yielded
 meaningful and interesting results [11]. The Einstein [3], Landau-Lifshitz
[4], Papapetrou [5], Bergmann-Thomson [6], Weinberg [7] and
Qadir-Sharif [8] definitions are coordinate dependent and the
calculations have to be done in Cartesian coordinates. Only the
M\o ller [9] prescription allows to perform the calculations in
any coordinate system. We also notice the similarity of some
results obtained with the energy-momentum complexes [3]-[9] with
the results given by their tele-parallel versions [12]. A great
contribution to the rehabilitation of the pseudotensors has been
done by Chang, Nester and Chen [13], they demonstrated that
different quasi-local definitions correspond to different boundary
conditions.\\
\\
 Recently, study of noncommutative geometry has emerged. To quantize the spacetime
 in string/M theory, it is realized that
 coordinates may become noncommutative operators on a D-brane[14] - [15] . The result
 is a discretization of spacetime where the spacetime
 coordinate operators satisfy the relation $[x^\mu,x^\nu ] = i \theta^{\mu
 \nu}$, where $\theta^{\mu
 \nu}$ is an anti symmetric matrix which determines the
 fundamental discretization of spacetime. It is shown that the
 divergences that appear  in General Relativity could be avoided
 if non commutativity replaces point like structures by smeared
 objects. The smearing effect is mathematically implemented with a
 substitution of Dirac delta function by a Gaussian distribution
 of minimal length $\sqrt{\theta}$. Schwarzschild spacetime is changed
 some what when a noncommutative spacetime is taken into account.
\\
\\
\\
 In this paper, we calculate the energy-momentum for a
non-commutative radiating Schwarzschild black hole [14]-[15] and
study some limiting cases. For our purpose, we use the Einstein
and M\o ller prescriptions.
\\
\\
The structure of our article is as follows: in Section II we
present the non-commutative radiating Schwarzschild black hole
[14]-[15]. In Section III,  we present the Einstein and M\o ller
energy-momentum complexes whereas in Section IV we performed the
calculations of the energy distributions  for the non-commutative
radiating Schwarzschild black hole. In Section V we briefly
present our concluding remarks. Throughout our work we use for
performing the calculations the signature ($1,-1,-1,-1$) and the
geometrized units ($c=1;G=1$). Also, Greek (Latin) indices take
value from $0 $ to $3$ and $1$ to $3$, respectively. \pagebreak

\section{Non-commutative Radiating Schwarzschild Black Hole}

In this section we present the non-commutative radiating Schwarzschild black
hole [14]-15] that is under study. The spacetime is described by the metric
given by
\\

$ds^{2}=\left[1-\frac{4M}{r\sqrt{\pi }}\gamma \left(\frac{3}{2},\frac{r^{2}}{4\theta }%
\right)\right]dt^{2}\\~~~~~~~~~~~~~~~~~~~~~~~~~-\frac{dr^{2}}{\left[1-\frac{4M}{r\sqrt{\pi }}\gamma \left(\frac{3}{2},\frac{%
r^{2}}{4\theta }\right)\right]}dr^{2}-r^{2} d\Omega^{2} ,$
\begin{equation} \tag{1}\end{equation}

where $\gamma $ is the lower incomplete gamma function that has
the expression $\gamma (\frac{3}{2},\frac{r^{2}}{4\theta })\equiv
\int\limits_{0}^{\frac{r^{2}}{4\theta }}\sqrt{t}\exp (-t)dt $. "In
flat spacetime noncommutativity eliminates point-like structures
in favor of smeared objects" [14]. The authors of [14] considered
the mass density of a static, spherically symmetric, smeared,
particle-like gravitational source given by

\begin{equation}
\rho _{\theta }(r)=\frac{M}{(4\pi \theta )^{3/2}}\exp
\left(-\frac{r^{2}}{4\theta }\right).  \tag{2}
\end{equation}

The particle of mass $M$ is not localized at a point, but is
diffused throughout a region of linear size $\sqrt{\theta }$. This
is the results of the intrinsic uncertainty that is encoded in the
coordinate commutator.

 At presently accessible energies, i.e. $\sqrt{\theta
}<10^{-16}$ cm the noncommutativity is not visible. We notice that
minimal deviations from standard vacuum Schwarzschild black hole
are expected at large distances. Also, at the distance $r\simeq
\sqrt{\theta }$ some behaviour of new physics is expected, because
in this case the density of energy and momentum is non negligible
and present. For balancing the inward gravitational pull and to
prevent droplet to collapse into a matter point the radial
pressure $p_{r}=-\rho _{\theta }$ has to be different by zero. The
spacetime non commutativity produces this important physical
effect on matter. Also, this implies the existence of the new
physics at the distance $r\simeq \sqrt{\theta }$.

 The Einstein equations
were solved considering $\rho _{\theta }(r)$ as a matter source and the
resulting gravitational background is given by (1).

The mass distribution is

\begin{equation}
m(r)\equiv \frac{2M}{\sqrt{\pi }}\gamma
\left(\frac{3}{2},\frac{r^{2}}{4\theta }\right), \tag{3}
\end{equation}

with $M$ being the total mass of the source. Analogous to the General
Relativity we have

\begin{equation}
m^{\prime }(r)=4\pi r^{2}\rho _{\theta }(r).  \tag{4}
\end{equation}

In the limit $r/\sqrt{\theta }\rightarrow \infty $ the classical
Schwarzschild black hole solution is recovered. The metric (1) can give"
useful insights about possible noncommutative effects on Hawking radiation"
[14].

\section{Einstein and M\o ller Energy-Momentum Complexes}

In this section we present the Einstein and M\o ller energy-momentum
complexes.

The Einstein energy-momentum complex [3] in a four-dimensional gravitational
background is given by
\begin{equation}
\theta _{\nu }^{\mu }=\frac{1}{16\pi }h_{\nu ,\,\lambda }^{\mu \lambda }.
\tag{5}
\end{equation}%
The Einstein superpotentials $h_{\nu }^{\mu \lambda }$ have the expression%
\begin{equation}
h_{\nu }^{\mu \lambda }=\frac{1}{\sqrt{-g}}g_{\nu \sigma }[-g(g^{\mu \sigma
}g^{\lambda \kappa }-g^{\lambda \sigma }g^{\mu \kappa })]_{,\kappa }  \tag{6}
\end{equation}%
and obey the antisymmetry property
\begin{equation}
h_{\nu }^{\mu \lambda }=-h_{\nu }^{\lambda \mu }.  \tag{7}
\end{equation}%
$\theta _{0}^0$ and $\theta _{i}^0$ represent the energy and
momentum density components, respectively. The Einstein
energy-momentum complex observes the local conservation law
\begin{equation}
\theta _{\nu ,\,\mu }^{\mu }=0.  \tag{8}
\end{equation}%
The energy and momentum in Einstein's definition are given by
\begin{equation}
P_{\mu }=\int \int \int \theta _{\mu }^{0}\,dx^{1}dx^{2}dx^{3}  \tag{9}
\end{equation}

and applying Gauss' theorem the energy-momentum is

\begin{equation}
P_{\mu }=\frac{1}{16\pi }\int \int h_{\mu }^{0i }n_{i}dS, \tag{10}
\end{equation}

where $n_{i}$ represents the outward unit normal vector over the surface $%
dS. $ Here $P_{0}=E$ is the energy.

In (9) and (10) $P_{i}$, $i=1,2,3$, represent the momentum
components.

The definition of the M{\o }ller energy-momentum complex [9] is given by

\begin{equation}
\mathcal{J}_{\nu }^{\mu }=\frac{1}{8\pi }M_{\nu \,\,,\,\lambda }^{\mu
\lambda },  \tag{12}
\end{equation}%
where we have the M{\o }ller superpotentials $M_{\nu }^{\mu \lambda }$ given
as below
\begin{equation}
M_{\nu }^{\mu \lambda }=\sqrt{-g}\left( \frac{\partial g_{\nu \sigma }}{%
\partial x^{\kappa }}-\frac{\partial g_{\nu \kappa }}{\partial x^{\sigma }}%
\right) g^{\mu \kappa }g^{\lambda \sigma }.  \tag{13}
\end{equation}%
The M{\o }ller superpotentials $M_{\nu }^{\mu \lambda }$ present the
antisymmetric property
\begin{equation}
M_{\nu }^{\mu \lambda }=-M_{\nu }^{\lambda \mu }.  \tag{14}
\end{equation}

Very important is that M{\o }ller's energy-momentum complex observes the
local conservation law
\begin{equation}
\frac{\partial \mathcal{J}_{\nu }^{\mu }}{\partial x^{\mu }}=0,  \tag{15}
\end{equation}%
where $\mathcal{J}_{0}^{0}$ is the energy density and $\mathcal{J}_{i}^{0}$
represents the momentum density components.

In the M\o ller definition the energy and momentum are given by
\begin{equation}
P_{\mu }=\int \int \int \mathcal{J}_{\mu }^{0}dx^{1}dx^{2}dx^{3}.  \tag{16}
\end{equation}%
\qquad \qquad The energy distribution is obtained with the expression
\begin{equation}
E=\int \int \int \mathcal{J}_{0}^{0}dx^{1}dx^{2}dx^{3}.  \tag{17}
\end{equation}%
\qquad With the aid of Gauss' theorem we obtain

\begin{equation}
P_{\mu }=\frac{1}{8\pi }\int \int M_{\mu }^{0i}n_{i}dS. \tag{18}
\end{equation}

In their important works Cooperstock [16] and Lessner [17]
stressed the importance of the M\o ller energy-momentum complex.
In addition, we notice the good results obtained with the Einstein
and M\o ller definitions for the energy-momentum in the case of
various geometries [11].

\section{Energy Distribution of the Non-commutative Radiating Schwarzschild
Black Hole}

The Einstein definition required Cartesian coordinates for
performing the calculations. We transform the gravitational
background given by (1) in Schwarzschild Cartesian coordinates, as
given by
\\

$ds^{2}=B(r)dt^{2}-(dx^{2}+dy^{2}+dz^{2})\\~~~~~~~~~~~~~~~~~~~~~~~~~-
\left[\frac{A(r)-1}{r^{2}}\right] (xdx+ydy+zdz)^{2}$,

\begin{equation}  \tag{19}  \end{equation}
with $B(r)=1-\frac{4M}{r\sqrt{\pi }}\gamma
(\frac{3}{2},\frac{r^{2}}{4\theta
})$ and $A(r)=\frac{1}{1-\frac{4M}{r\sqrt{\pi }}\gamma (\frac{3}{2},\frac{%
r^{2}}{4\theta })}$.\\

 We use Maple program with the GRTensor II
attached package to calculate the energy distribution and momenta
and to make plots.

The Einstein superpotentials that we use for the evaluation of the
energy distribution $h_{\nu }^{\mu \lambda }$ are given by

\begin{equation}
h_{0}^{0x}=\frac{2x}{r^{2}}\frac{4M}{r\sqrt{\pi }}\gamma \left(\frac{3}{2},\frac{%
r^{2}}{4\theta }\right),  \tag{20}
\end{equation}

\begin{equation}
h_{0}^{0y}=\frac{2y}{r^{2}}\frac{4M}{r\sqrt{\pi }}\gamma \left(\frac{3}{2},\frac{%
r^{2}}{4\theta }\right),  \tag{21}
\end{equation}

\begin{equation}
h_{0}^{0z}=\frac{2z}{r^{2}}\frac{4M}{r\sqrt{\pi }}\gamma \left(\frac{3}{2},\frac{%
r^{2}}{4\theta }\right).  \tag{22}
\end{equation}

Using (10) and (20)-(22) the expression for the energy
distribution in the Einstein definition is given

\begin{equation}
E_{E}=M-\frac{Mr}{\sqrt{\pi }\sqrt{\theta }}\exp
\left(-\frac{r^{2}}{4\theta }\right)-M
 erf c\left(\frac{1}{2}\frac{r}{\sqrt{\theta }}\right).  \tag{23}
\end{equation}

The energy depends on the mass $M$, $\theta $ parameter and radial coordinate. In the limit case $r/\sqrt{%
\theta }\rightarrow \infty $ we obtain the energy of the classical
Schwarzschild black hole solution $E_{E}=M$.

The M{\o }ller superpotential involved in the calculation of the energy $%
M_{0}^{0t}$ is

$M_{0}^{0t}=[2M-\frac{2Mr}{\sqrt{\pi }\sqrt{\theta }}\exp
(-\frac{r^{2}}{ 4\theta })-\frac{Mr^{3}}{\sqrt{\theta }\theta
\sqrt{\pi }}\exp (-\frac{r^{2} }{4\theta })
\\
\\~~~~~~~~~~~~~~~~~~~~~~~~~~~~-2M
{erf}c(\frac{1}{2}\frac{r}{\sqrt{\theta }})]\sin \theta$.
\begin{equation} \tag{24}
\end{equation}
The energy distribution in the M\o ller prescription is obtained
combining (18) with (24)
\\

$E_{M}=M-\frac{Mr}{\sqrt{\pi }\sqrt{\theta }}\exp \left(-\frac{r^{2}}{4\theta }\right)-%
\frac{Mr^{3}}{2\sqrt{\theta }\theta \sqrt{\pi }}\exp
\left(-\frac{r^{2}}{4\theta }\right)\\
~~~~~~~~~~~~~~~~~~~~~~~~~~~~~~ - M erf
c\left(\frac{1}{2}\frac{r}{\sqrt{\theta }}\right).$
    \begin{equation}   \tag{25}  \end{equation}

In this case, the energy distribution also presents a dependence on the mass
$M$, $\theta $ parameter and radial coordinate. In the M\o ller definition also for the limit $r/\sqrt{%
\theta }\rightarrow \infty $, we recovered the energy of the
classical Schwarzschild black hole solution $E_{M}=M.$

\begin{figure}
\vspace{0.4cm} \includegraphics[width=0.2\textwidth]{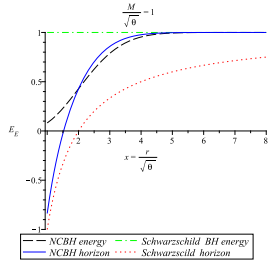}
\caption{The plot for the energy E prescribed by Einstein vs.
$x=\frac{r}{\sqrt{\theta}}$. Solid  and dotted curves cut the x at
the horizons of NCBH and Schwarzschild black hole.} \label{fig:1}
\end{figure}
\begin{figure}
\vspace{0.4cm} \includegraphics[width=0.2\textwidth]{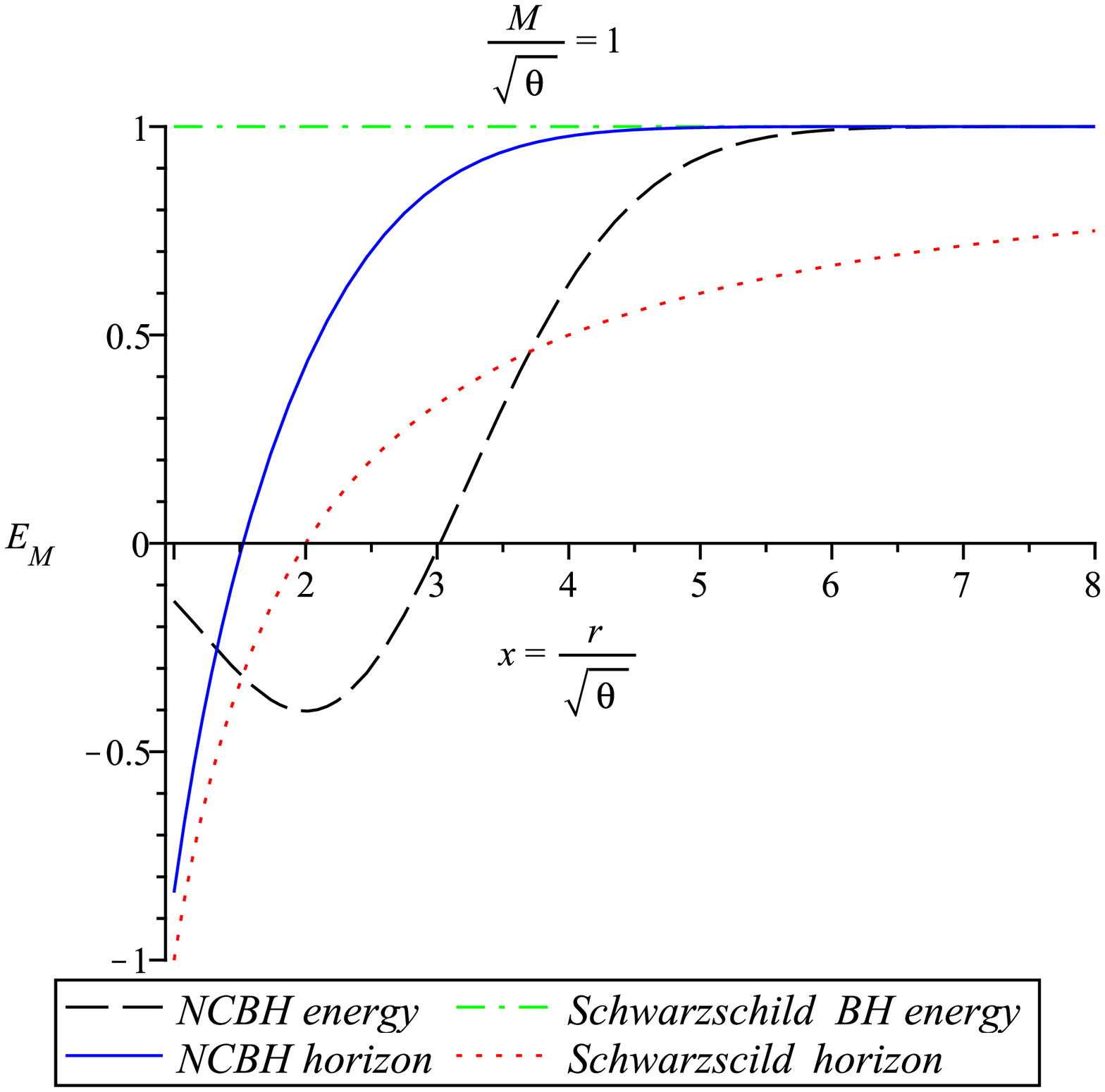}
\caption{The plot for the energy E prescribed by M\o ller vs.
$x=\frac{r}{\sqrt{\theta}}$. Solid  and dotted curves cut the x at
the horizons of NCBH and Schwarzschild black hole.} \label{fig:1}
\end{figure}
\begin{figure}
\vspace{0.4cm} \includegraphics[width=0.2\textwidth]{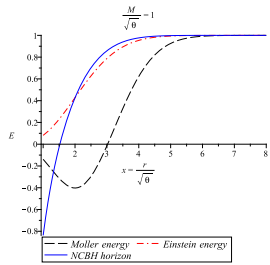}
\caption{Comparison of the  energy E prescribed by M\o ller and
Einstein vs. $x=\frac{r}{\sqrt{\theta}}$. Solid curve cuts the x
at the horizon.} \label{fig:1}
\end{figure}
\begin{figure}
\vspace{0.4cm} \includegraphics[width=0.3\textwidth]{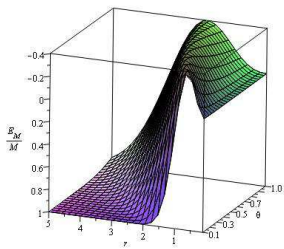}
\caption{Variation of the  energy E prescribed by M\o ller with
respect to r and $\theta$.} \label{fig:1}
\end{figure}
\begin{figure}
\vspace{0.4cm} \includegraphics[width=0.3\textwidth]{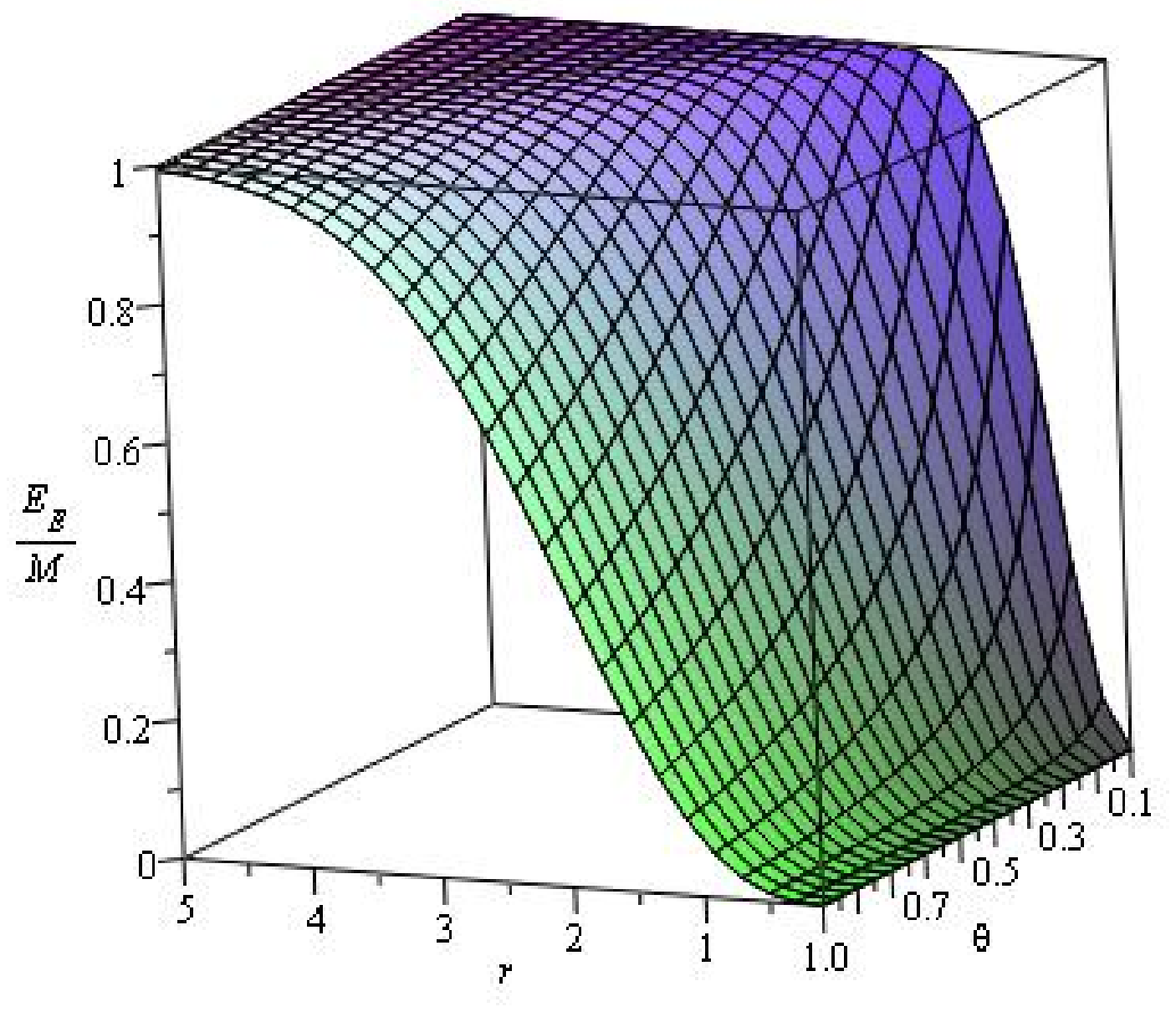}
\caption{Variation of the  energy E prescribed by Einstein with
respect to r and $\theta$.} \label{fig:1}
\end{figure}
\begin{figure}
\vspace{0.4cm} \includegraphics[width=0.2\textwidth]{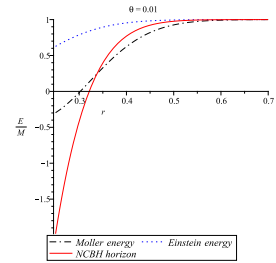}
\caption{Variation of the  energy E prescribed by Einstein and
 M\o ller with respect to r for a fixed $\theta =.01$.} \label{fig:1}
\end{figure}

\pagebreak

\section{Final Remarks}

In our paper, we study the energy distribution of a
non-commutative radiating Schwarzschild black hole in the Einstein
and M\o ller prescriptions. In both
prescriptions the expressions for energy depend on the mass $M$%
, $\theta $ parameter as well as radial coordinate. One can
observe from the figures 1-3 that the energy in Einstein's
prescription is always positive whereas the energy in M\o ller's
prescription assumes positive  values only after a certain
distance from the horizon. Interestingly, we note that the energy
in Einstein's prescription adopts some real values within the
horizon. However, in the limiting case $r/\sqrt{\theta
}\rightarrow \infty $ both yield the same expression for energy as
$E_{E}=E_{M}=M$ that corresponds to the case of the classical
Schwarzschild black hole solution.  This also represents the ADM
mass. The above limit can be achieved in two ways: either, $
\theta \rightarrow 0$ i.e. when the noncommutativity is not
visible or $r \rightarrow \infty $ i.e. at large distance. When
$r\approx \sqrt{\theta} $, then $E_M \neq E_E$. In fact $E_E
>0 $
 and $E_M<0$. Our results show that the Einstein prescription is a powerful concept
  than M\o ller's prescription. This is also sustained by the meaningful results obtained with the Einstein definition. One can mention the
 work of Virbhadra [18], where he emphasized the importance of the Einstein prescription.
\pagebreak

 \large{\textbf{Acknowledgements}:}
\\
\\
 FR is  thankful to  UGC, Govt. of India,  for ~~~providing financial support under Research Award scheme.
\\
\\
\\
\\
\\
 \large{

\end{document}